# Microwave Transport in Metallic Single-Walled Carbon Nanotubes

Z. Yu, P.J. Burke
*Integrated Nanosystems Research Facility, Department of Electrical Engineering and Computer Science,
University of California, Irvine, CA 92697-2625*

The dynamical conductance of electrically contacted single-walled carbon nanotubes is measured from dc to 10 GHz as a function of source-drain voltage in both the low-field and high-field limits. The ac conductance of the nanotube itself is found to be equal to the dc conductance over the frequency range studied for tubes in both the ballistic and diffusive limit. This clearly demonstrates that nanotubes can carry high-frequency currents at least as well as dc currents over a wide range of operating conditions. Although a detailed theoretical explanation is still lacking, we present a phenomenological model of the ac impedance of a carbon nanotube in the presence of scattering that is consistent with these results.

There are two reasons to study single-walled carbon nanotubes (SWNT)[1]: physics and technology. For physics, single walled nanotubes represent nearly ideal 1d electronic structures which allow for experimental studies of interaction effects and Luttinger liquid behavior[2], ballistic transport at room temperature[3], quantum confinement effects, Coulomb blockade at room temperature[4], and spin transport[5]. For technology, nanotube transistors are predicted to be extremely fast[6], especially if the nanotubes can be used as the interconnects themselves in future *integrated nanosystems*. The extremely high mobilities found in semiconducting nanowires[7] and nanotubes[8] are important for high speed operations, one of the main predicted advantages of nanotube and nanowire devices in general[9]. Nanotubes may also have a role to play as high frequency interconnects in the long term between active nanotube transistors or in the short term between conventional transistors because of their capacity for large current densities.

Early theoretical work[10], as well as our recent circuit modeling work[11], predict significant frequency dependence in the nanotube dynamical impedance in the absence of scattering and contact resistance. The origin of this predicted frequency dependence is in the collective motion of the electrons, which can be thought of as 1d plasmons. Our equivalent circuit description shows that the nanotube forms a quantum transmission line, with distributed kinetic inductance and both quantum and geometric capacitance. (One of us recently verified the 2d analog of this effect[12].) In the absence of damping, standing waves on this transmission line can give rise to resonant frequencies in the microwave range (1-10 GHz) for nanotube lengths between 10 and 100 μm. We also proposed an ad-hoc damping model, relating the damping to the dc resistance per unit length. To date, there have been no measurements of the microwave frequency conductance of a SWNT to either confirm or deny these theoretical predictions or equivalent circuit models.

In this Letter, we present the first measurements of the high frequency conductance of a single walled nanotube. We find experimentally that the ac conductance is equal to the dc conductance up to at least 10 GHz. This clearly demonstrates for the first time that the current carrying capacity of carbon nanotubes can be extended without degradation into the high frequency (microwave) regime.

In our experimental results, no clear signatures of Tomonaga-Luttinger liquid behavior are observed (in the form of non-trivial frequency dependence) and no specifically quantum effects (reflecting quantum versus classical conductance of nanotubes) are reported, in contradiction to theoretical predictions for ac conductance in 1d systems that neglect scattering[10]. In order to explain this discrepancy between theory (which neglects scattering) and experiment (which includes realistic scattering), we present a phenomenological model for the finite frequency conductance of a carbon nanotube which treats scattering as a distributed resistance. This model explains why our results at ac frequencies do not display frequency dependence. Simply put, resistive damping washes out the predicted frequency dependence.

Individual SWNTs[13] were synthesized via chemical vapor deposition[14,15] on oxidized, high-resistivity p-doped Si wafers ($\rho > 10$ kΩ-cm) with a 400-500 nm SiO$_2$ layer. Metal electrodes were formed on the SWNTs using electron-beam lithography and metal evaporation of 20-nm Cr/100 nm Au bilayer. The devices were not annealed. Nanotubes with electrode spacing of 1 (device A) and 25 μm (device B) were studied. Typical resistances were ~ MΩ; some nanotubes had resistances below 250 kΩ. In this study we focus on metallic SWNTs (defined by absence of a gate response) with resistance below 200 kΩ. Measurements were performed at room temperature in air.

Fig. 1 shows the room temperature I-V characteristic of device A, a SWNT with a 1 μm electrode spacing. Since this length is comparable to the mean-free-path, this device is in the quasi-ballistic limit. The low-bias resistance of this device was 60 kΩ. This resistance is most likely dominantly due to the contact; at low fields, once electrons are injected transport is quasi-ballistic from source to drain. The device clearly shows saturation in the current at around 20 μA. The inset shows that (over almost the entire range of applied voltage) the absolute resistance (V/I) can be described by a simple function

$$V/I = R_0 + |V|/I_0, \qquad (1)$$

where $R_0$ and $I_0$ are constants, as was originally found and explained by Yao[16]. From the slope of the linear part of the R-V curve, we find $I_0 = 29$ μA for this device, in good agreement with Yao[16]. There, it was shown that the saturation behavior is due to a modified mean-free-path for electrons when the electric field is sufficient to accelerate electrons to a large enough energy to emit an optical phonon. This effect was studied more quantitatively with similar conclusions in [17,18].

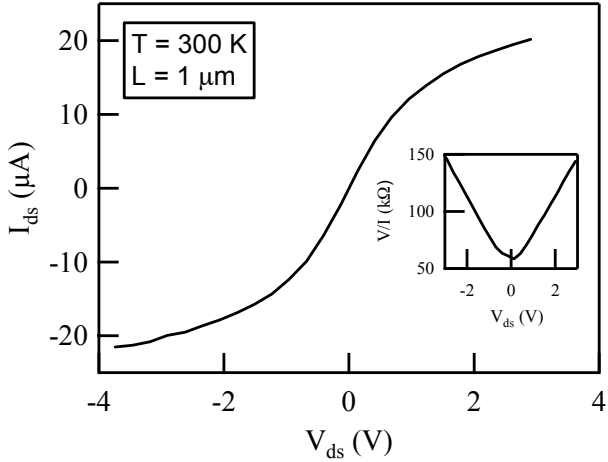

**Figure 1**. Current-voltage characteristic for device A.

In order to measure the dynamical impedance at microwave frequencies, a commercially available microwave probe (suitable for calibration with a commercially available open/short/load calibration standard) allowed for transition from coax to lithographically fabricated on chip electrodes. The electrode geometry consisted of two small contact pads, one 50x50 μm$^2$, and the other 200x200 μm$^2$ (for device A) or 50x200 μm$^2$ (for device B). A microwave network analyzer is used to measure the calibrated (complex) reflection coefficient $S_{11}(\omega) \equiv V_{reflected}/V_{incident}$, where $V_{incident}$ is the amplitude of the incident microwave signal on the coax, and similarly for $V_{reflected}$. This is related to the load impedance $Z(\omega)$ by the usual reflection formula: $S_{11}=[Z(\omega)-50\ \Omega]/[Z(\omega)+50\ \Omega]$. At the power levels used (3 μW), the results are independent of the power used.

The statistical error in the measurement of both the Re($S_{11}$) and Im($S_{11}$) due to random noise in the network analyzer is less than 1 part in $10^4$. A systematic source of error in the measurement due to contact-to-contact variation and non-idealities in the calibration standard gives rise to an error of 2 parts in $10^3$ in the measurement of Re($S_{11}$) and Im($S_{11}$). Because the nanotube impedance is so large compared to 50 Ω, these errors will be important, as we discuss in more depth below.

We measure the value of $S_{11}$ as a function of frequency and source-drain voltage for both device A and B. *While the absolute value of $S_{11}$ is found to be $0 \pm 0.02$ dB over the frequency range studied (the systematic error due to contact-to-contact variation), small changes in $S_{11}$ with the source-drain voltage are systematic, reproducible, and well-resolved within the statistical error of $\pm 0.0005$ dB. The change in $S_{11}$ with source-drain voltage is not an artifact, since control samples do not exhibit this effect.* Our measurement clearly shows that the value of $S_{11}$, and hence the nanotube dynamical impedance, depends on the dc source-drain bias voltage, and that this dependence is independent of frequency over the range studied for both devices.

For both device A and B, we find Im($S_{11}$)=0.000 ± 0.002, indicating that the nanotube impedance itself is dominantly real. Our measurement system is not sensitive to imaginary impedances much smaller than the real impedance, which is of order 100 kΩ. For all measurements presented here, Im($S_{11}$) does not change with $V_{ds}$ within the statistical uncertainty of 1 part in $10^4$. On the other hand, Re($S_{11}$) changes reproducibly with $V_{ds}$, indicating that the real part of the nanotube dynamical impedance changes with $V_{ds}$.

By linearizing the relationship between $S_{11}$ and the conductance G, it can be shown that for small values of G (compared to 50 Ω), G(mS) ≈ 1.1 x $S_{11}$(dB). (We note that after calibration, a control experiment with no nanotube gives 0 ± 0.02 dB, where the uncertainty is due to variations in the probe location on the contact pads from contact to contact.) Based on this calculation, we conclude that the absolute value of the measured high frequency conductance is found to be 0 with an error of ± 22 μS, which is consistent with the dc conductance.

In order to analyze the data more quantitatively, we concentrate on the *change* in $S_{11}$ with $V_{ds}$. The measurement error on the *change* in the ac conductance G with bias voltage depends primarily on the statistical uncertainty in $S_{11}$, which in our experiments is 20 times lower than the systematic error. (Since the contact probe remains fixed in place while changing the gate voltage, we can reproducibly and reliable measure small changes in $S_{11}$ with the source-drain voltage.) Thus, although the absolute value of G can only be measured with an uncertainty of 20 μS, a change in G can be measured with an uncertainty of 1 μS. These uncertainties are a general feature of any broadband microwave measurement system.

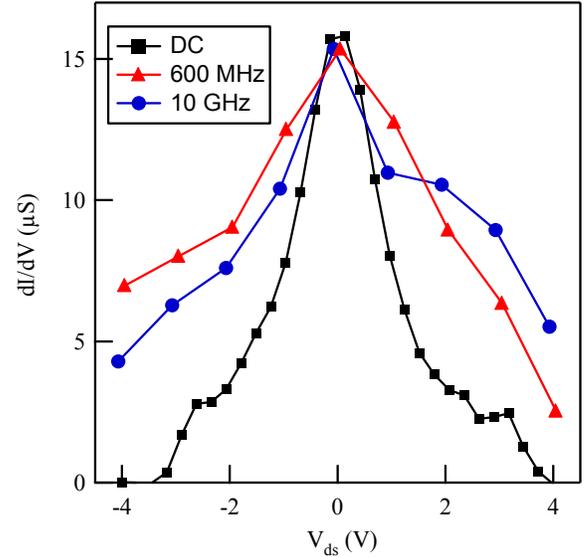

**Figure 2**. Conductance vs. $V_{ds}$ for device A at DC, 0.6 GHz, 10 GHz.

In Fig. 2 we plot G vs. the source-drain voltage at dc, 0.6 GHz, and 10 GHz for device A. We only know the change in G with $V_{ds}$, so we add an offset to $G_{ac}$ to equal $G_{dc}$ at $V_{ds}$=0.

We discuss this in more detail below, but at the moment it is clear that the G at ac changes with $V_{ds}$ just as it does at dc. We now discuss the offset.

Based on the measured results we know the absolute value of G is between 0 and 22 µS; based on Fig. 2 we know that G changes by 10 µS when $V_{ds}$ changes by 4 V. The dynamical conductance is probably not negative (there is no physical reason for this to be the case), which allows the following argument to be made: Since $G_{ac}(V_{ds}=0)-G_{ac}(V_{ds}=4V)=10$ µS (measured), and $G_{ac}(V_{ds}=4V) > 0$ (on physical grounds), therefore $G_{ac}(V_{ds}=0V)>10$ µS; our measurements put this as a lower limit; the upper limit would be 20 µS. Therefore, our measurements show for the first time that, within 50%, nanotubes can carry microwave currents just as efficiently as dc currents.

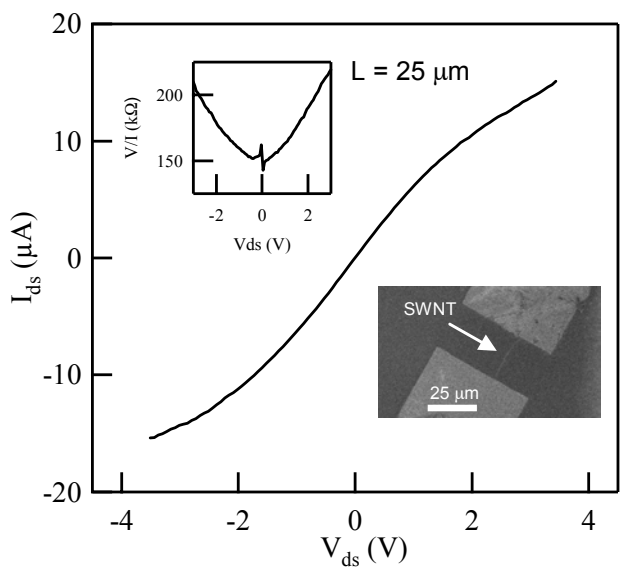

**Figure 3**. I-V curve and SEM (inset) of device B.

Because device A is in the quasi-ballistic limit, but does not approach the theoretical lower limit of 6 kΩ for perfect contacts, the metal-nanotube contact resistance probably dominates the total resistance for this sample. In order to focus more heavily on the nanotube resistance itself, we turn now to device B.

In Fig.3, we plot the I-V curve of a longer SWNT (device B), with an electrode gap of 25 µm. (The original length of this nanotube was over 200 µm.) This device is almost certainly not in the ballistic limit, even for low-bias conduction, since the mean-free-path is of order 1 µm[15,17,18] and the SWNT length is 25 µm. The low-bias resistance of this device is 150 kΩ. Previous measurements in our lab[15] on 4 mm long SWNTs gave a resistance per unit length of 6 kΩ/µm, indicating that the SWNT bulk resistance is about 150 kΩ for device B, and that the contact resistance is small compared to the intrinsic nanotube resistance. The absolute resistance (V/I) and the source-drain I-V curve for this device is well-described by Eq. 1, as for device A. We find $I_0=34$ µA for this device, in agreement with device A.

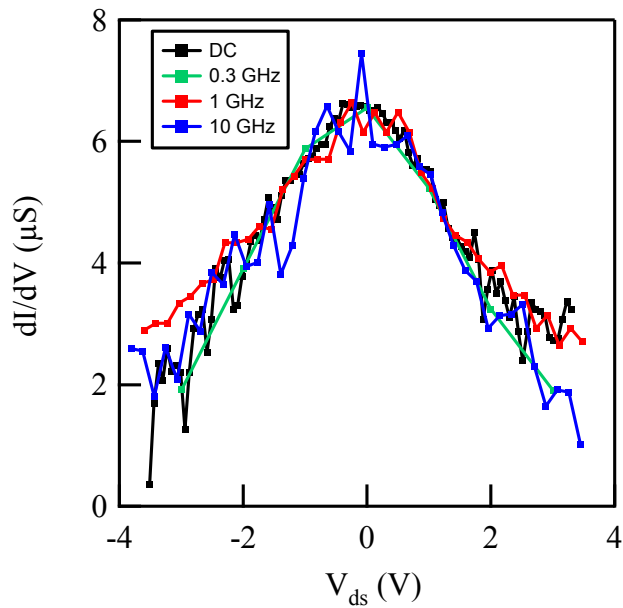

**Figure 4.** Conductance (AC and DC) for device B.

In Fig. 4 we plot G vs. the source-drain voltage at dc, 0.3 GHz, 1 GHz, and 10 GHz for device B. As for device A, we only know the change in G with $V_{ds}$, so we add an offset to $G_{ac}$ to equal $G_{dc}$ at $V_{ds}=0$. It is clear from this graph that the nanotube dynamical conductance changes with bias voltage just as the dc conductance does. Using similar arguments as for device A, our measurements for device B show that the ac and dc conductance are equal within 50% over the entire frequency range studied.

We now turn to a discussion of our results. At DC, the effects of scattering on nanotubes have been well-studied[16-18]. The dc resistance is given by[19]

$$R_{dc} = \frac{h}{4e^2} \frac{L_{nanotube}}{l_{m.f.p.}}, \qquad (2)$$

where $l_{m.f.p.}$ is the mean-free-path. In ballistic systems, the sample contact resistance dominates and the dc resistance has a lower limit given by $h/4e^2 = 6$ kΩ, which is possible only if electron injection from the electrodes is reflectionless. Is equation (2) true at finite frequencies? The answer to this question in general is not known.

For the simple case of an ohmically contacted nanotube of length L, we have predicted the first resonance would occur at a frequency given by $v_F/(4Lg)$, where $v_F$ is the Fermi velocity, L the length, and g the Luttinger liquid "g-factor", a parameter which characterizes the strength of the electron-electron interaction. Typically, g ~ 0.3. For L = 25 µm, the first resonance in the frequency dependent impedance would occur at 24 GHz, beyond the range of frequencies studied here. However, our nanotube for device B was originally over 200 µm long. After deposition of electrodes, the nanotube extended under the two electrodes for a distance of at least 150 μm on one side, and 50 μm on the other. If these segments of the nanotube were intact, it would correspond to plasmon resonances at frequencies of 4 and 8 GHz. We clearly do not observe any strong resonant behavior at these or any other frequencies. We believe this must be due to the damping of these plasmons, as we discuss below.

While this is not justified rigorously, we assume that equation (2) describes a distributed resistance of the nanotube that is independent of frequency, equal to the measured dc resistance per unit length of 6 kΩ/μm of similar long nanotubes grown in our lab[15]. In our previous modeling work[11], we found that (under such heavy damping conditions) the nanotube dynamical impedance is predicted to be equal to its dc resistance for frequencies less than $1/(2\pi R_{dc} C_{total})$, where $C_{total}$ is the total capacitance of the nanotube (quantum and electrostatic). Although our measurements presented here are on top of a poorly conducting ground plane (high resistivity Si), and the previous modeling work was for a highly conducting substrate, we can use the modeling as a qualitative guide. For device B, we estimate $C_{total}$=1 fF, so that the ac impedance would be predicted to be equal to the dc resistance for frequencies below about ~ 1 GHz. This is qualitatively consistent with what we observe experimentally.

At high bias voltages, the electrons have enough energy to emit optical phonons, dramatically reducing the mean-free-path and modifying equation (2) to the more general equation (1). Our measurements clearly show that equation (1) is still valid up to 10 GHz. A theoretical explanation for this is lacking at this time, although it is intuitively to be expected for the following reason: the electron-phonon scattering frequency in the high-bias region is approximately 1 THz[18]. Therefore, on the time-scale of the electric field period, the scattering frequency is instantaneous. Further theoretical work is needed to clarify this point.

Measurements up to higher frequencies of order the electron-phonon scattering rate (~ 50 GHz at low electric fields[18]) should allow more information to be learned about electron-phonon scattering in nanotubes; temperature dependent measurements would allow for more information as well, such as the intrinsic nanotube impedance at low scattering rates.

Acknowledgements: This work was supported by the Army Research Office (award DAAD19-02-1-0387), the Office of the Naval Research (award N00014-02-1-0456), and DARPA (award N66001-03-1-8914), and the National Science Foundation (award ECS-0300557, CCF-0403582).